\documentstyle[amssymb,aps,multicol,epsfig]{revtex}

\begin{document}
\title{Quantum magneto-oscillations in a two-dimensional Fermi liquid}
\author{Gregory W. Martin, Dmitrii L. Maslov, and Michael Yu. Reizer\cite{reizer}}
\address{Department of Physics, University of Florida, P.O. Box 118440, Gainesville, FL 32611}
\date{\today}
\maketitle
\begin{abstract}
Quantum magneto-oscillations provide a powerfull tool for quantifying
Fermi-liquid parameters of metals. In 
particular, the quasiparticle effective mass and spin susceptibility are 
extracted from the experiment using the 
Lifshitz-Kosevich formula, derived under the assumption that the properties of the 
system in a non-zero magnetic field are determined uniquely by the zero-field
Fermi-liquid state. This 
assumption is valid in 3D but, generally speaking, erroneous in 2D where the Lifshitz-Kosevich formula may be applied only 
if the oscillations are strongly damped by thermal smearing and disorder. 
In this work,  the effects of interactions and disorder on the amplitude of 
magneto-oscillations in 2D are studied. It is found that the effective mass diverges 
logarithmically with decreasing temperature signaling a deviation from the Fermi-liquid behavior. 
It is also shown that the quasiparticle lifetime due to inelastic interactions does not enter the oscillation 
amplitude, although these interactions do renormalize the effective mass. This result provides 
a generalization of the Fowler-Prange theorem  formulated originally for the electron-phonon 
interaction.
\end{abstract}

\begin{multicols}{2}
Patterns of quantum magneto-oscillations in thermodynamic (de Haas-van Alphen effect)
and transport (Shubnikov-de Haas effect) quantities encode three important
parameters of a Fermi-liquid metal. The period of the oscillations gives the
area of the extremal cross-section of the Fermi surface, the slope of the
temperature dependence of the oscillations amplitude provides the
quasiparticle effective mass, and the phase shift between
oscillations of spin-up and-down electrons yields the (renormalized) spin
susceptibility. Magneto-oscillations studies of the Fermi-liquid (FL) state 
in two-dimensions (2D) date back
to the early 70s, when semiconductor heterostructures first became
available \cite{ando}. Another surge of the activity in this field, which
occurred in mid 90s, was stimulated by the discovery of the metallic state
at $\nu =1/2$ \cite{compositefermions}. Recently, Shubnikov-de Haas
oscillations have been used to determine the parameters of the
``anomalous'' metallic state in Si MOSFETs and other semiconductor
heterostructures exhibiting an apparent metal-insulator transition in zero 
magnetic field \cite{shayegan,vitkalov,kravchenko,pudalov1}.
Despite the long and successful history of quantifying FLs in 3D
via magneto-oscillations, this method remains controversial in 2D. The
primary goal of our paper is to resolve some of the open issues.

The first controversy is related to the applicability of the current theory
of magneto-oscillations to the 2D case. The analysis of the experimental data in 2D is often
based on the premise that the classic result for magneto-oscillations in a
Fermi liquid for the 3D case, known
as the ``Lifshitz-Kosevich (LK) formula'' \cite{lifshitz,luttinger,bychkov}, is 
transferrable to 2D upon a trivial change in the
electron spectrum. The crucial features of the LK formula, i.e., its
validity for arbitrarily strong interactions (without destroying the FL) and 
the fact that the FL parameters
entering the formula are taken at {\em zero }magnetic field,
survive on this premise. The deviations of the observed oscillation pattern in stronger fields 
from that predicted by the LK formula are ascribed to oscillations in the effective
$g$-factor \cite{ando} and the effective mass \cite{smith}.
On the other hand, there have been warnings
that the 3D LK formula is non-transferrable to 2D 
\cite{varma,stamp} for any field strength. Hence the situation needs to be clarified. The 
second controversy--not specific to 2D--is related to the effect of quasiparticle
damping. It is often mentioned in the literature that any scattering of
quasiparticles, elastic and inelastic, contributes to the smearing of
magneto-oscillations via the effective Dingle temperature (scattering rate)
for a given process. Alternatively, Fowler and Prange \cite{fowler}
showed that the electron-phonon scattering rate does not
appear in the oscillations amplitude due to the cancellation
of two $T-$dependent parts of the Matsubara self-energy (cf. also \cite
{engelsberg}). To the best of our knowledge, this cancellation has never
been discussed for other interactions, including the
electron-electron one, which is of a primary importance for 2D electron
systems. The last issue to be addressed in this paper
(and not discussed previously in the literature) is the
effect of interference between electron-impurity and electron-electron
scattering on magneto-oscillations, neglected in the LK formula.
The theory of interference effects in the ballistic regime \cite{ZNA}, when $T\tau \gg 1,$ where 
$\tau$ is the electron-impurity scattering time (we set $\hbar=k_B=1$ throughout
the paper), offers a plausible explanation of the metallic temperature
dependence in the metallic phase of the 2D metal-insulator transition.
Unusual (within the LK framework) temperature dependences of the oscillation 
amplitude are also commonly observed in Si MOSFETs \cite{pudalov1,pudalov_private}, 
but the proper theory is currently lacking. 

Our answers to these open questions are as follows.
i) Although it is true that the LK formula does not work in 2D
at $T=0$ and in the absence of disorder, it is still applicable to the
situation when finite temperature and/or disorder cause 
the oscillations to be exponentially small. ii) The cancellation of the
scattering rate term in the Matsubara self-energy is pertinent to {\em any} inelastic
interaction, including the electron-electron one. Due to this cancellation, 
the scattering rate of inelastic processes does not enter the oscillation amplitude. 
iii) Interference between electron-impurity and electron-electron
interactions gives a new, $T\ln T$, dependence of the
amplitude's argument, that can be interpreted equivalently either as a 
``$T-$dependent'' effective mass or Dingle temperature. The
functional form of this dependence is the {\em same }in the diffusive 
$\left( T\tau \ll 1\right)$ and ballistic $\left( T\tau \gg 1\right)$ regimes.

We limit our
analysis to the de Haas-van Alphen effect and assume the chemical
potential is fixed. The main features of the results for the de Haas-van Alphen
effect, in particular, the $T$-dependence of the oscillation amplitude, are commonly
expected to apply to the Shubnikov-de Haas effect as well, although a rigorous proof
of that is currently lacking. Assuming a fixed chemical potential is
not essential for the case of small oscillations, which is the focus
of this paper (see below). We begin with a brief reminder of how the LK formula is derived in the
Luttinger formalism \cite{luttinger,bychkov}.
The key issue here is whether the zero-field FL parameters determine
uniquely the oscillation pattern in a finite (and not small) field. The
(Matsubara) self-energy (that encodes all FL parameters) has consists of
parts $\Sigma=\Sigma_{0}+\Sigma_{\text{osc}},$ where $\Sigma_{0}$ 
may contain a monotonic (nonoscillatory) dependence on
magnetic field and ${\Sigma}_{\text{osc}}$ oscillates with the
field. For electron-electron interactions in 3D, 
$|\Sigma_{\text{osc}}|/|\Sigma_{0}|\sim N^{-3/2}$, where $N$ is the number of occupied Landau levels, 
whereas the leading term in the oscillatory part of the thermodynamic potential $\Omega $ 
falls off as $N^{-5/2}$ for $\Sigma_{\text{osc}}=0$. Expanding $\Omega$ in ${\Sigma}_{\text{osc}}$ 
up to the second order (the first order term
vanishes due to the property $\delta \Omega /\delta {\Sigma}=0$), 
one finds that the oscillatory part of ${\Sigma}$ can always be 
neglected in the semiclassical regime ($N\gg 1$). With this simplification 
and for a momentum-independent self-energy, the amplitude of the $k^{\text{th}}$ harmonic 
in $\Omega$ is given by 
\begin{equation}
A_{k}=\frac{4\pi ^{2}kT}{\omega _{c}}\sum_{\varepsilon _{n}>0}\exp
\left( -\frac{2\pi k\left[ \varepsilon _{n}+i\Sigma_{0}\left(i
\varepsilon _{n},T\right) \right] }{\omega _{c}}\right),  \label{ampl1}
\end{equation}
where $\varepsilon_n=\pi(2n+1)T$ and $\omega _{c}=eB/mc$. Notice that 
$i\Sigma_{0}$ is real and
does not contain a constant term.
The second argument of $\Sigma_{0}$ emphasizes the fact that the
temperature enters $\Sigma_{0}$ in two ways: via the Matsubara frequency
and via the thermal distribution of electrons and other degrees of freedom. 
For a generic Fermi liquid and in the presence of short-range impurities, 
$i\Sigma_{0}\left(i \varepsilon _{n},T\right) =\alpha \varepsilon
_{n}+\text{sgn}\varepsilon _{n}/2\tau$,
so that the effective mass is defined as $m^{\ast }=m(1+\alpha )$. The amplitude 
then assumes a familiar form 
\begin{equation}
A_{k}=\frac{2\pi ^{2}kT/\omega _{c}}{\sinh \left( 2\pi ^{2}kT/\omega
_{c}^{\ast }\right) }\exp \left( -\frac{2\pi ^{2}T_{D}}{\omega _{c}}\right),
\label{ampl2}
\end{equation}
where $\omega_c^*=eB/m^*c$ and $T_{D}=1/2\pi \tau$ is the Dingle temperature. 
Momentum-dependence of $\Sigma_{0}$ of the form 
$\beta v_{F}\left( p-p_{F}\right)$ results in a change of the
effective mass in (\ref{ampl2}) to 
$m^{\ast }=m\left( 1+\alpha \right) (1+\beta )^{-1}$ and in multiplying (\ref{ampl2}) 
by $Z_{s}m^{\ast }/m$, where $Z_{s}$ is the renormalization factor. 

In arbitrary dimensionality $D$, the estimates for the ratio of oscillatory
to monotonic-in-field parts of the self-energy and for the leading
oscillatory term in $\Omega$ change to 
$N^{-D/2}$ and $N^{-(D+2)/2}$, respectively. For $D=2$,
the oscillations in the self-energy are as important as in the thermodynamic
potential itself \cite{stamp}. The Luttinger expansion at $T=T_{D}=0$ breaks 
down and the LK formula is not, generally speaking, valid \cite{stamp}. The 
physical reason is that the ground states of an interacting system at $B=0$ 
and in a finite field are not adiabatically connected in 2D.
This fact has been emphasized by
recent findings that the ground state of a 2D electron liquid is not a Fermi
liquid even for $N\gg 1$, but rather a charge-ordered
state \cite{stripes}. Nevertheless, an absense of the full LK formula in 2D
does not preclude a canonical analysis of magneto-oscillations, if under
more restrictive conditions, as the FL-behavior is restored at higher
energies.

The power-counting argument for (against) neglecting the oscillatory part of
the self-energy in 3D (2D) is valid at $T=T_{D}=0$. If the real and/or Dingle 
temperatures are sufficiently high, i.e., 
\begin{equation}
2\pi ^{2}\left( T/\omega _{c}^{\ast }+T_{D}/\omega _{c}\right) \gtrsim 1,
\label{condition}
\end{equation}
the amplitudes of {\em all} oscillatory quantities, including the
self-energy, are exponentially small. 
Neglecting the oscillatory part of the self-energy, the amplitude of the
first harmonic takes the form 
\begin{equation}
A_{1}=\left( 4\pi ^{2}T/\omega _{c}\right) \exp \left( -2\pi \left[ \pi T+i%
\Sigma_{0}\left(i \pi T,T\right) \right] /\omega _{c}\right) ,
\label{ampl3}
\end{equation}
where, due to condition (\ref{condition}), we limited the Matsubara sum by
the first term $\varepsilon _{0}=\pi T$. The oscillatory part of $\Sigma$ results in a
correction to $A_{1}$ which is itself of order $A_{1}$ (with exponential
accuracy). The net contribution to $\Omega$ is of order $A_{1}^{2}$, which is 
of the same order as $A_{2}$ for $\Sigma_{\text{osc}}=0$. Thus, harmonics 
with $k\geq 2$ are affected by the
oscillations in $\Sigma$ and a 2D analog of the LK formula, which
includes the sum over all $k$, can only be derived in a perturbation theory 
for a specific interaction but not for a generic FL. However, the $k=1$ harmonic does not
include $\Sigma_{\text{osc}}$ and, as long as (\ref{condition}) is satisfied, 
the analysis can proceed as in the 3D case. 
In what follows, we assume that (\ref{condition}) is satisfied and the
amplitude of the first (and only important) harmonic is given by (\ref {ampl3}).

Next we discuss whether the quasiparticle relaxation rate affects the
amplitude of magneto-oscillations. We set $T_{D}=0$ temporarily. 
Suppose that a quasiparticle relaxation rate is
measured in a clean Fermi liquid, e.g., via electron heat conductivity, with a result that 
$1/\tau _{\text{e-e}}\propto T^{2}$.
It seems natural to assume that the same rate 
contributes also to the Dingle temperature of magneto-oscillations. That this is 
{\em not} the case was shown for the electron-phonon interaction by Fowler and
Prange \cite{fowler}. Here we generalize their arguments for the
electron-electron interaction in 3D and 2D, and then give a general theorem
for an arbitrary interaction. For a generic FL in 3D, the Matsubara
self-energy, up to the quadratic in $\varepsilon _{n}$ and $T$ terms, can be
written as 
\begin{equation}
\!\!\!\!i\Sigma_{0}\left(i \varepsilon _{n},T\right) =\alpha \varepsilon
_{n}+i\beta v_{F}(p-p_{F})+\gamma \left[ \left( \pi T\right)
^{2}-\varepsilon _{n}^{2}\right] .  \label{sigma3Dquad}
\end{equation}
In addition to a direct calculation, the validity of the quadratic term 
in (\ref{sigma3Dquad}) is readily established by noticing that upon analytic continuation 
$i\varepsilon_{n}\rightarrow \varepsilon +i0$ this term gives the correct form for the
imaginary part of the on-shell self-energy: $-%
\mathop{\rm Im}%
\Sigma _{0}^{R}\propto \left( \pi T\right) ^{2}+\varepsilon ^{2}$ \cite{AGD}. 
The amplitude of the first harmonic (\ref{ampl3}) contains 
$i\Sigma_{0}\left(i \pi T,T\right)$, in which the quadratic term vanishes identically.
The $T ^{2}$-terms from higher Matsubara frequencies (legitimately considered 
within this scheme in 3D) increase the amplitude and cannot be interpreted as "damping."

In 2D, the integral over momentum transfers diverges logarithmically at the
lower limit, changing the behavior of $%
\mathop{\rm Im}%
\Sigma _{0}^{R}$ to $E^{2}\ln E$, where $E=\mbox{max}\{\varepsilon,T\}$. 
This change does not alter the principal result. Consider the simplest case 
of a contact interaction. To the second order in this interaction, the quadratic term in (\ref{sigma3Dquad}) is replaced by 
\begin{equation}
i\tilde{\Sigma}_{0}(i \varepsilon _{n},T)=-\frac{U^{2}m}{\pi ^{2}v_F^2}T\sum_{\omega
_{m}=0}^{\varepsilon _{n}-\pi T}\omega _{m}\ln \left(\varepsilon_F/\omega
_{m}\right)\label{sum}. 
\end{equation}
Although the sum in (\ref{sum}) does not have an analytic solution, it obviously
vanishes for $\varepsilon _{n}=\pi T$. 

To analyze a general case of a finite-range and dynamic interaction,
including the screened Coulomb one, it is convenient to find the
imaginary part of the retarded self-energy first and then continue back to
Matsubara frequencies. On the mass-shell, $%
\mathop{\rm Im}%
\Sigma _{0}^{R}$ is given by 
\[
\mathop{\rm Im}%
\Sigma^{R}_0\left( \varepsilon \right) =-\frac{\pi }
{2\left( 2\pi \right) ^{D}}\int d\omega F\left( \omega \right) \left[ \coth \frac{\omega}{2T}%
-\tanh\frac{\omega -\varepsilon}{2T} \right], 
\]
where 
$F\left( \omega \right) =\int d^{D}q\delta
\left( \omega -{\bf v}_{F}\cdot {\bf q}\right) 
\mathop{\rm Im}%
V^{R}\left( \omega ,q\right)$ and $V^{R}\left( \omega ,q\right)$ is the retarded interaction potential.
As a function of a complex variable $z$, $f(z)\equiv%
\mathop{\rm Im}%
\Sigma ^{R}\left( z\right) $ has the following properties in the upper
half-plane: i) all lines $%
\mathop{\rm Im}%
z=\pi \left( 2n+1\right) T$ are branch cuts on which $%
\mathop{\rm Re}%
f$ is continuous but $%
\mathop{\rm Im}%
f$ changes jumpwise; ii) due to the fact that $\tanh(x-i\pi(n+1/2))=\coth x$%
, all points $z=i\pi \left( 2n+1\right) T$ are zeroes of $f(z).$ Thus,
function $f(z)$ is analytic in the band $0\leq 
\mathop{\rm Im}%
z<\pi T$ including the point $z=i\pi T$. Analytic continuation from the real
axis into this band is legitimate and at $z=i\pi T$ it yields the Matsubara
self-energy $\tilde{\Sigma}_{0}(i \varepsilon _{0}=i \pi T,T)$, which is
equal to zero. Zeroes of $f\left( z\right)$ at $z=i\pi \left( 2n+1\right) T$
with $n\geq 1$ do not lead to vanishing of 
$\tilde{\Sigma}_{0}(i \varepsilon _{n},T)$ for $n\geq 1$
because those zeroes are separated from the real axis by branch cuts and
thus are not accessible by analytic continuation. The 2D case is special
only in that the $q-$integration results in the $\ln \omega$ factor in
$F(\omega)$ which does not change the reasoning given above. In particular, 
for a dynamically screened Coulomb interaction in 2D, 
$F\left( \omega \right) \propto \omega \ln |\omega|$ and still 
$\tilde{\Sigma}_{0}(i \pi T,T)=0$. As this result does 
not depend on the particular form of the interaction, it can be viewed as a 
generalization of the Fowler-Prange theorem. Notice that in 3D the Fowler-Prange
theorem is of limited applicability because nothing prevents one from considering
$k>1$ and lower values of $T+T_D$, when the effect of $\Sigma_0(i \varepsilon_{n>0},T)$
needs to be taken into account. In 2D, one is bound to consider only 
$\Sigma_0(i \varepsilon_{n=0},T)$ within the Luttinger approximation.

\begin{figure}[tbp]
\centerline{\epsfxsize=3.0in \epsfbox{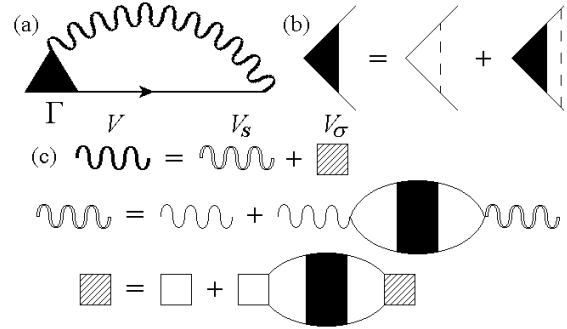}}
\caption{(a) The interference correction to the self-energy. (b) The vertex correction 
is assigned to either one of the vertices in (a) because the self-energy arises as 
insertions into the thermodynamic potential (closed loops). (c) Singlet/triplet-channel 
contributions to the effective potential.}
\end{figure}

Finally, we discuss the effect of interference between electron-electron and
electron-impurity scattering on magneto-oscillations, extending the analysis of the interference
corrections to the self-energy in 2D from the diffusive ($T\tau \ll 1$) \cite{altshuler} 
to the ballistic ($T\tau\gg 1$) limit. The general form of the interference correction to the
Matsubara self-energy is (see Fig. 1) 
\begin{eqnarray}
\Sigma_{0}^{\text{int}}\left(i \varepsilon _{n}, {\bf p}\right) &=&-2T%
\sum _{ \varepsilon_{n} \left( \omega _{m}- \varepsilon _{n}\right)>0}%
\int \frac{d^{2}q}{\left( 2\pi \right) ^{2}} V(i \omega_{m},q)%
\label{sigmainter} \\
\!&&\times \Gamma\left(i \omega _{m},q\right)%
G \left( i \varepsilon _{n}-i \omega _{m}, {\bf p-q}\right) ,  \nonumber
\end{eqnarray}
where $G \left( i \varepsilon _{n}, {\bf p}\right) =\left(i \varepsilon _{n}-\xi _{p}+i\text{sgn}%
\varepsilon _{n}/2\tau \right) ^{-1}$ and the effective interaction 
$V=V_{s}+3V_{t}$ contains the contributions from both singlet and
triplet channels \cite{ZNA}
\begin{eqnarray}
V_{s}^{-1}\left(i \omega _{m},q\right) &=&(2\pi e^2/q
+F_{\rho }^{0}/\nu )^{-1}-{\Pi}\left(i \omega _{m},q\right);\nonumber\\
V_{\sigma }^{-1}\left(i \omega _{m},q\right) &=&\nu /F_{\sigma }^{0}-%
{\Pi}\left(i \omega _{m},q\right),\nonumber
\end{eqnarray}
where $\nu=m/\pi$. The factor of two in (\ref{sigmainter}) accounts for
two possibilities of including the vertex correction in Fig.1a. The Fermi-liquid 
constants $F_{\rho }^{0}$ and $F_{\sigma }^{0}$ \cite{FLconstants} determine the renormalized 
charge and spin-susceptibilities, respectively. The vertex 
$\Gamma\left(i \omega _{m},q\right) =
(\sqrt{\left( \left| \omega
_{m}\right| \tau +1\right) ^{2}+\left( qv_F\tau\right) ^{2}}-1) ^{-1}
$
reduces to $\Gamma=\left( D\tau q^{2}+\left| \omega _{m}\right| \tau
\right) ^{-1}$ and $\Gamma=\tau ^{-1}\left( \left| \omega _{m}\right|
^{2}+\left( qv_{F}\right) ^{2}\right) ^{-1/2}$ in the diffusive and
ballistic limits, respectively. The general form of the
(small $q$) polarizarion operator 
\begin{eqnarray}
{\Pi}\left(i \omega _{m},q\right) =-\nu \left[ 1-\left| \omega
_{m}\right| \tau \Gamma\left(i \omega _{m},q\right) \right] \label{P}
\end{eqnarray}
reduces to ${\Pi}\left(i \omega _{m},q\right) =-\nu
Dq^{2}/\left( Dq^{2}+\left| \omega _{m}\right| \right) $ in the diffusive
limit, where $D=v_{F}^{2}\tau /2$, and to ${\Pi }^{0}(i \omega _{m}, q)=-\nu
\left( 1-\left|\omega_{m}\right|/\sqrt{\left( v_{F}q\right) ^{2}+\omega _{m}^{2}}\right)$
in the ballistic one.  Omitting the details of lengthy but straightforward calculations, 
we give just the result for the self-energy valid to logarithmic accuracy 
\[
i\Sigma_{0}^{\text{int}}\left(i \pi T,T\right) =- T\ln\left(\varepsilon_F/T\right) 
Q\left( T\tau ,F_{\sigma}^{0}\right)/2\varepsilon _{F}\tau,
\]
where $Q\left( T\tau ,F_{\sigma }^{0}\right) = g_{\rho }\left( T\tau \right) +
\left[3F_{\sigma }^{0}/\left(1+F_{\sigma }^{0}\right)\right]g_{\sigma }\left(T\tau \right)$
and $g_{\rho /\sigma }\left( x\right) $ are slowly varying functions which
interpolate between the diffusive and ballistic regimes. The limiting values
of $g_{\rho /\sigma }\left( x\right) $ are as follows: $g_{\rho }\left(
0\right) = 1,$ $g_{\rho }\left( x\gg 1\right) = 3/2,$ $g_{\sigma }\left(
0\right) = 1,$ $g_{\sigma }\left( x\gg 1\right) = 1/2$. Apart from the 
numerical coefficients, the $T-$dependence of $i\Sigma_{0}^{\text{int}}\left( \pi T,T\right)$ 
is the same in the diffusive and ballistic regimes. In that sense, the behavior
of the self-energy is sumilar to that of the tunneling density of states 
\cite{rudin}. Notice that the interference correction 
to the {\em scattering rate} in the ballistic regime $\left| 
\mathop{\rm Im}%
\Sigma _{0}^{R}\right| _{\text{int}}\sim \left( T/\varepsilon _{F}\tau
\right) \ln \left( \varepsilon _{F}/T\right)$ is smaller than the
scattering rate in a clean FL, $\left| 
\mathop{\rm Im}%
\Sigma _{0}^{R}\right| _{\text{int}}\sim \left( T^{2}/\varepsilon
_{F}\right) \ln \left( \varepsilon _{F}/T\right)$, in the parameter $\left(
T\tau \right) ^{-1}\ll 1$. However, due to the cancellation of the $T^{2}\ln
T$ term in $i\Sigma_{0}\left(i \pi T,T\right)$ discussed earlier in this
paper, the interference correction is the main nonlinear $T-$dependent
term in the Matsubara self-energy, leading to a modification of the
LK-formula. The $T\ln T-$dependence of the self-energy can be interpreted
as a logarithmic $T-$dependence of the effective mass. Following this
interpretation, the interference effect leads to a replacement of the 
effective mass in the argument of the exponential in $A_1$ by 
\begin{equation}
m^{\ast }\left( T\right) =m^{\ast }\left( 1-\frac{m}{m^{\ast }}\frac{\ln
\left( \varepsilon _{F}/T\right) }{2\pi \varepsilon _{F}\tau }Q\left( T\tau
,F_{\sigma }^{0}\right) \right).  \label{mstarT}
\end{equation}
Notice that the effective mass is reduced by the singlet-channel interaction
but enhanced by the ferromagnetic $\left( F_{\sigma }^{0}<0\right) $
interaction in the triplet channel. The $\ln T-$dependence of the effective
mass is a characteristic feature of the ``marginal Fermi liquid'' model \cite
{MFL,peltzer}.

Equivalently, the nonlinear $T-$dependence of 
$i\Sigma_{0}\left(i \pi T,T\right)$ may be interpreted as a $T-$dependent Dingle temperature: 
\begin{equation}
T_{D}\left( T\right) =T_{D}\left( 1-\left(T/\varepsilon _{F}\right)
\ln\left
(\varepsilon _{F}/T\right)Q\left( T\tau ,F_{\sigma }^{0}\right) \right) .
\label{TDT}
\end{equation}
One of the empirical procedures used in \cite{pudalov1} to account for
the observed deviations from the LK formula was to assume that the
effective Dingle temperature has the same $T-$dependence as the zero-field
resistivity. Comparing (\ref{TDT}) with the result for the interference
correction to the resistivity \cite{ZNA}, we see that, although this recipe
is not precise, it has some theoretical justification:\ the general structure of
the $T_{D}\left( T\right)$ and $\rho \left( T\right)$ depedences is
similar in the ballistic regime, except for a factor of $\ln \left(
\varepsilon _{F}/T\right)$ present in $T_{D}\left( T\right)$ but not in 
$\rho \left( T\right)$.

This work was supported by NSF DMR-0077825. We are grateful to I. L. Aleiner, A. V. Chubukov, J. Klauder, and
B. Z. Spivak for valuable comments and to M. E. Gershenson and V. M. Pudalov for
extensive discussions of the experimental data.

\end{multicols}

\begin{references}
\bibitem{FLconstants}  As in \cite {ZNA}, we make a model assumption of $F_{\sigma }^{0}$ to have 
the same value in the diffusive and ballistic limits.
\end{references}
\end{document}